\documentclass[aps,showpacs,twocolumn,prl]{revtex4}
\usepackage{epsfig, amssymb}

\begin{document}

\newcommand{\de}{\Delta E}
\newcommand{\mbc}{M_{bc}}
\newcommand{\pp}{p\bar{p}}
\newcommand{\bb}{B\bar{B}}
\newcommand{\kpi}{K^-\pi^+}
\newcommand{\kpipi}{K^-\pi^+\pi^+}
\newcommand{\kpipin}{K^-\pi^+\pi^0}
\newcommand{\kpipipi}{K^-\pi^+\pi^+\pi^-}
\newcommand{\dkpi}{D^0\to\kpi}
\newcommand{\dkpipin}{D^0\to\kpipin}
\newcommand{\dkpipipi}{D^0\to\kpipipi}
\newcommand{\dpkpipi}{D^+\to\kpipi}
\newcommand{\dsdpin}{D^{*0}\to D^0\pi^0}
\newcommand{\dspdpi}{D^{*+}\to D^0\pi^+}
\newcommand{\bdpp}{B\to D \pp}
\newcommand{\bdnpp}{\bar{B^0}\to D^0\pp}
\newcommand{\bdppp}{B^+\to D^+\pp}
\newcommand{\bdspp}{B\to D^*\pp}
\newcommand{\bdsnpp}{\bar{B^0}\to D^{*0}\pp}
\newcommand{\bdsppp}{B^+\to D^{*+}\pp}
\newcommand{\bppk}{B^+\to \pp K^+}

\title{\Large \rm 
Observation of $\bar{B^0}\to D^{(*)0}\pp$}

\author{
  K.~Abe$^{9}$,               % KEK
  K.~Abe$^{44}$,              % TohokuGakuin
% N.~Abe$^{47}$,              % TIT
  R.~Abe$^{30}$,              % Niigata
  T.~Abe$^{45}$,              % Tohoku
% I.~Adachi$^{9}$,            % KEK
  Byoung~Sup~Ahn$^{16}$,      % Korea
  H.~Aihara$^{46}$,           % Tokyo
  M.~Akatsu$^{23}$,           % Nagoya
% M.~Asai$^{10}$,             % Hiroshima
  Y.~Asano$^{51}$,            % Tsukuba
  T.~Aso$^{50}$,              % Toyama
  V.~Aulchenko$^{2}$
% ...
\\         % Tsukuba
\vspace*{10pt}
(Belle Collaboration)\\
\small
\vspace*{10pt}
% ...
$^{2}${Budker Institute of Nuclear Physics, Novosibirsk}\\
% ...
$^{9}${High Energy Accelerator Research Organization (KEK), Tsukuba}\\
% ...
$^{16}${Korea University, Seoul}\\
% ...
$^{23}${Nagoya University, Nagoya}\\
% ...
$^{30}${Niigata University, Niigata}\\
% ...
$^{44}${Tohoku Gakuin University, Tagajo}\\
$^{45}${Tohoku University, Sendai}\\
$^{46}${University of Tokyo, Tokyo}\\
% ...
$^{50}${Toyama National College of Maritime Technology, Toyama}\\
$^{51}${University of Tsukuba, Tsukuba}\\
% ...
\normalsize
}

\begin{abstract}
The $B$ meson decay modes $\bdpp$ and $\bdspp$ have been studied
using 29.4 fb$^{-1}$ of data collected with the Belle detector at KEKB.
The $\bdnpp$ and $\bdsnpp$ decays have been observed for the first
time with branching fractions
${\cal{B}}({\bdnpp}) =(1.18\pm 0.15\pm 0.16)\times10^{-4}$ and
${\cal{B}}(\bdsnpp) =(1.20^{+0.33}_{-0.29}\pm 0.21)\times 10^{-4}$.
No signal has been found for the $\bdppp$ and $\bdsppp$ 
decay modes, and the corresponding upper limits at 90\% C.L. are 
presented.
\end{abstract}

\maketitle

\clearpage
{\ }
\clearpage

%\section{Introduction}
To date, information on $B$ decays with baryons in the final states 
is rather scarce.
A recent search for the two-body baryonic $B$ decays by 
Belle showed that their relative probabilities are rather small:
for the decay modes $B^0\to\pp,\  \Lambda{\bar\Lambda}$, and 
$B^+ \to p{\bar\Lambda}$ upper limits of $(1-2)\times 10^{-6}$ 
were obtained~\cite{belle_baryon}.
At the same time, measurements of the $B^0\to D^{*-}\pp\pi^+$ and 
$B^0\to D^{*-}p\bar{n}$ decay branching fractions by 
CLEO~\cite{dstar_p_nbar} and the observation of the $\bppk$ decay
by Belle~\cite{bkpp} indicate the dominance of
multibody final states in decays of $B$ mesons into baryons.
Moreover, the observation of the $\bar{B}^0\to D^0\pi^0$, 
$\bar{B}^0\to D^0\eta$ and $\bar{B}^0\to D^0\omega$ decays by 
Belle~\cite{dpi0_belle} and CLEO~\cite{dpi0_cleo} with branching
fractions considerably higher than expected indicates 
that color-suppressed $B$ decays with baryons in the
final state are likely to be sizeable.
This motivated a search for the color-suppressed decays 
$B \to D^{(*)}\pp$.
The inclusion of charge conjugate modes is implicit throughout this
report.

We use a data sample collected with the Belle detector
at the KEKB asymmetric energy $e^+e^-$ collider~\cite{KEKB}.
It consists of 29.4~fb$^{-1}$ taken at the $\Upsilon(4S)$ resonance
corresponding to $N_{\bb}=31.9\times 10^6$ produced $\bb$ pairs, and 
2.3~fb$^{-1}$ taken 60 MeV below the $\Upsilon(4S)$ resonance
to perform systematic studies of the $e^+e^-\to q\bar q$ background.

%\section{The Belle detector}
The Belle detector~\cite{NIM} is a large-solid-angle magnetic
spectrometer that consists of a three-layer silicon vertex
detector (SVD), a 50-layer central drift chamber (CDC) for charged
particle tracking and specific ionization measurement ($dE/dx$),
an array of aerogel threshold \v{C}erenkov counters (ACC),
time-of-flight scintillation counters (TOF), and a CsI(Tl) 
electromagnetic calorimeter (ECL) located in the magnetic volume.
The magnetic field is returned via an iron yoke that is instrumented
to detect muons and $K_L$ mesons (KLM). 

%\section{Event Selection}
Charged tracks are selected with requirements based on the 
average hit residual and impact parameter relative to the 
interaction point. We also require that the transverse momentum of 
the tracks be greater than 0.1 GeV/c to reduce the low momentum 
combinatorial background. 

For particle identification (PID), the combined information
from CDC, TOF and ACC subsystems is used.
Protons and antiprotons are selected with a set of PID criteria that has
an efficiency of 98\% and a kaon misidentification probability of 15\%.
Selection criteria for charged kaons provide
an efficiency of 88\%, a pion misidentification probability of 8\%, and
negligible contamination from protons.
All tracks positively identified as electrons are rejected.

A pair of calorimeter showers with an invariant mass within 
15~MeV/c$^2$ of the nominal $\pi^0$ mass is considered as a $\pi^0$ 
candidate. 
An energy of at least 50 MeV and a photon-like shape are required for 
each shower.

We reconstruct $D$ mesons in the following decay channels:
$\dkpi$, $\dkpipipi$, $\dkpipin$ and $\dpkpipi$.
We select $D$ candidates using cuts around the central
values of the $M(D)$ distributions that correspond to $95\%$ efficiency.
For the $\pi^0$ from the $\dkpipin$ decay, we require that 
the $\pi^0$ momentum in the $\Upsilon(4S)$ center-of-mass (CM) frame 
be greater than 0.2 GeV/c in order to reduce combinatorial background.
$D^*$ mesons are reconstructed in the $\dsdpin$ and $\dspdpi$ 
decay modes. Since the pions from $D^{*}\to D\pi$ decays are slow, we 
relax these cuts and impose an energy threshold for $\pi^0$ photons of 
30~MeV and a $\pi^{\pm}$ transverse momentum threshold of 50~MeV/c.
The mass difference between $D^*$ and $D$ candidates is required to be
within 4~MeV from the expected value for $D^{*0}$ and 2.5~MeV for
$D^{*+}$ ($\sim3\sigma$ in both cases).

We combine $D^{(*)}$ candidates with $\pp$ pairs to form $B$ mesons.
Candidate events are identified by their CM 
energy difference, \mbox{$\de=(\sum_iE_i)-E_b$}, and the
beam constrained mass, $\mbc=\sqrt{E_b^2-(\sum_i\vec{p}_i)^2}$, where 
$E_b$ is the beam energy and $\vec{p}_i$ and $E_i$ are the momenta and
energies of the decay products of the $B$ meson in the CM frame.
We select events with $\mbc>5.20$~GeV/c$^2$ and $|\de|<0.2$~GeV,
and define a $B$ signal region of
$5.272$~GeV/c$^2<\mbc<5.288$~GeV/c$^2$ and $|\de|<0.020$~GeV.
In the cases when there is more than one candidate in an event, 
the $\bdpp$ (or $\bdspp$) candidate with the $D$ mass (or $D^*-D$ 
mass difference) closest to the world average is chosen.
We use Monte Carlo (MC) simulation with a three-body phase space
distribution for the $B\to D^{(*)}\pp$ decays to model the response of
the detector and determine the efficiency~\cite{GEANT}.

%\section{Background Suppression}
To suppress the large combinatorial background that is dominated by 
the two-jet-like $e^+e^-\to~q\bar{q}$ continuum process, 
variables that characterize the event topology are used. We require 
$|\cos\theta_{\rm thr}|<0.80$, where $\theta_{\rm thr}$ is the angle 
between the thrust axis of the $B$ candidate and that of the rest of 
the event. % in the CM frame. 
This cut eliminates 77\% of the continuum background and 
retains 78\% of the signal events. We also define a Fisher discriminant, 
${\cal F}$, which includes: the production angle of the $B$ candidate, 
the angle of the $B$ candidate thrust axis with respect to the beam 
axis, and nine parameters that characterize the momentum flow in the 
event relative to the $B$ candidate thrust axis in the CM 
frame~\cite{VCal}. We impose a requirement on ${\cal{F}}$ that 
rejects 28\% of the remaining continuum background and retains 95\% of 
the signal.

%\section{Results of the analysis}
%\subsection{Calculation of branching fractions}
The $\de$ distributions were fitted with a Gaussian for the signal and
a linear function for the background. The Gaussian mean value and width 
were fixed from the MC simulation of the signal events.
In the fit to the $\de$ distribution, the region $\de<-0.1$~GeV is
excluded to avoid contributions from other $B$ decays, such as 
$B\to D^{(*)}\pi\pp$.
For the calculation of branching fractions, we use the signal yields
determined from the fit to the $\de$ distribution. This minimizes a 
possible bias from other $B$ meson decays, which tend to peak in $\mbc$ 
but not in $\de$.

\begin{figure}[t]
   \includegraphics[width=0.49\textwidth]{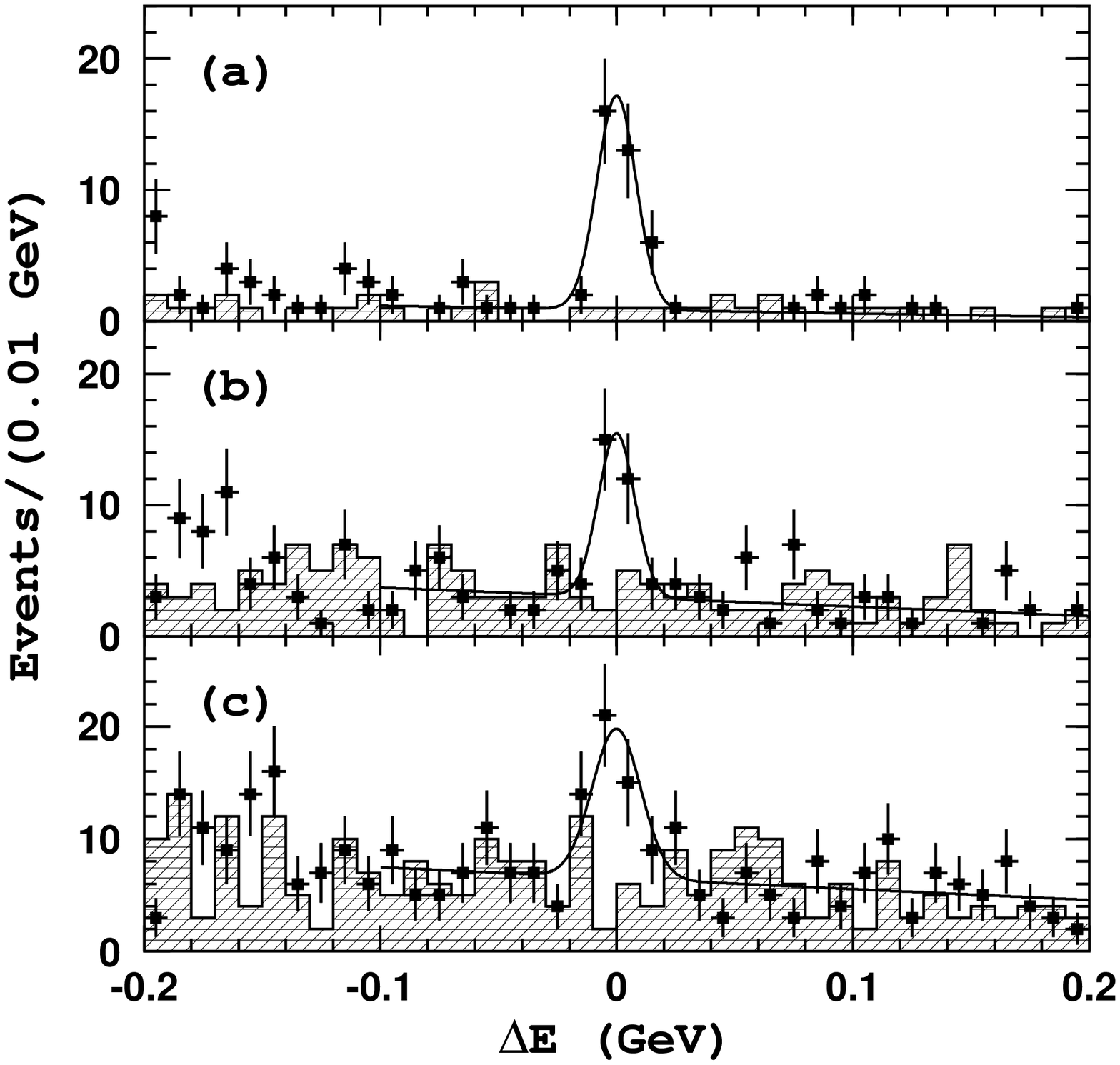}
  \caption{The $\de$ distributions for the $\bdnpp$ candidates:
  (a) $\dkpi$, (b) $\dkpipipi$ and (c) $\dkpipin$. 
  The points with errors are experimental data, the hatched histograms 
  are $D^0$ mass sidebands and the curves are fit results.}
  \label{d0_dembc}
\end{figure}
\begin{figure}
  \includegraphics[width=0.49\textwidth]{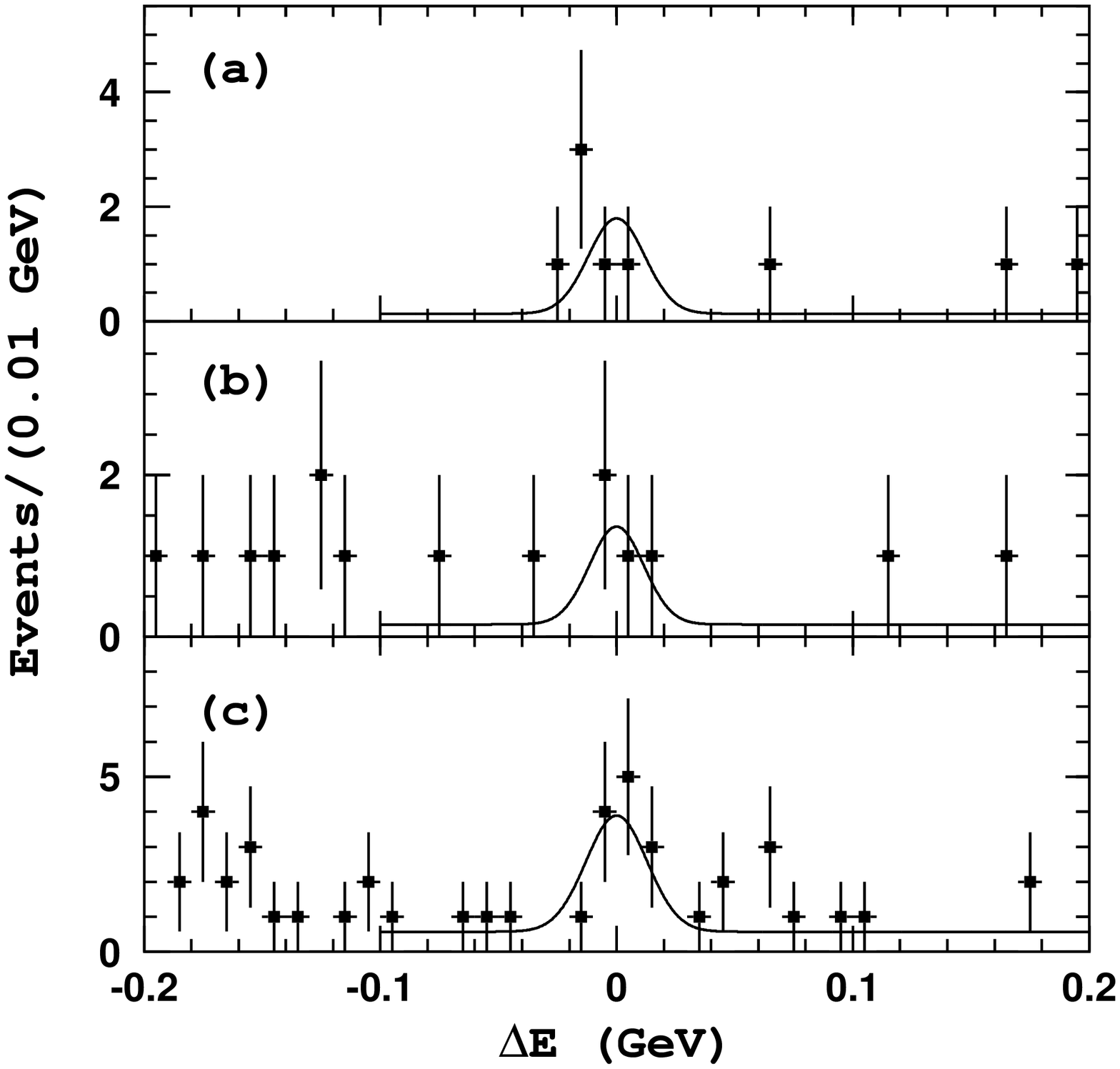}
  \caption{The $\de$ distributions for the $\bdsnpp$ candidates:
  (a) $\dkpi$, (b) $\dkpipipi$ and 
  (c) $\dkpipin$. The points with errors are experimental data 
  and the curves are fit results.}
  \label{ds0_dembc}
\end{figure}

The $\de$ distributions for the $\bdnpp$ and $\bdsnpp$ decays 
are shown in Figs.~\ref{d0_dembc} and \ref{ds0_dembc}. 
The fit results are presented in Table~\ref{defit}. 
Statistically significant signals are observed for the $\bdnpp$ decay 
mode in all three decay channels of the $D^0$ meson.
The corresponding branching fractions are in good agreement with 
each other.
\begin{table*}
\caption{Branching fractions and 90\% C.L. upper limits for
$B\to D^{(*)}\pp$ decays.}
\smallskip
\label{defit}
  \begin{tabular*}{\textwidth}{l@{\extracolsep{\fill}}ccccc}\hline\hline
Mode  & $\de$ yield & $\mbc$ yield & 
Efficiency, \% & ${\cal B}$, $10^{-4}$ & Significance\\\hline
%%%
$\bdnpp$, $\dkpi$ & $33.6^{+6.5}_{-5.8}$ & $34.5^{+6.5}_{-5.8}$ &
        $23.56\pm 0.49$ & $1.17^{+0.23}_{-0.20} \pm 0.14$ & $8.9\sigma$\\

$\bdnpp$, $\dkpipipi$ & $24.2^{+6.3}_{-5.7}$ & $14.7^{+5.8}_{-5.1}$ &
        $7.11\pm 0.21$ & $1.42^{+0.37}_{-0.34} \pm 0.22$ & $5.6\sigma$\\

$\bdnpp$, $\dkpipin$ &  $34.2^{+8.6}_{-7.9}$ & $36.5^{+8.2}_{-7.4}$ &
        $7.28\pm 0.28$ & $1.06^{+0.27}_{-0.24} \pm 0.15$ & $5.1\sigma$\\

$\bdnpp$, simultaneous fit & --- & --- & 
        --- & $1.18\pm 0.15 \pm 0.16$ & $12\sigma$\\\hline
%%%
$\bdsnpp$, $\dsdpin$, $\dkpi$ & 
        $5.0^{+2.8}_{-2.2}$ & $6.1^{+2.9}_{-2.3}$ &
        $8.11\pm 0.30$ & $0.81^{+0.46}_{-0.36} \pm 0.13$ & $2.9\sigma$\\

$\bdsnpp$, $\dsdpin$, $\dkpipipi$ & 
        $3.5^{+2.4}_{-1.7}$ & $2.6^{+2.4}_{-1.8}$ & 
        $1.96\pm 0.15$ & $1.21^{+0.82}_{-0.59} \pm 0.23$ & $2.6\sigma$\\

$\bdsnpp$, $\dsdpin$, $\dkpipin$ &
        $10.8^{+4.0}_{-3.4}$ & $13.6^{+4.4}_{-3.8}$ &   
        $2.38\pm 0.16$ & $1.65^{+0.61}_{-0.52} \pm 0.30$ & $4.2\sigma$\\

$\bdsnpp$, simultaneous fit & --- & --- & --- & 
        $1.20^{+0.33}_{-0.29}\pm 0.21$ & $5.6\sigma$\\\hline
%%%%
$\bdppp$, $\dpkpipi$ & $<5.2$ & $<5.1$ & 
        $14.28\pm 0.38$ & $<0.15$ 90\% C.L. & ---\\\hline

$\bdsppp$, $\dspdpi$, $\dkpi$ & $<2.2$ & $<2.3$ &
        $9.55\pm 0.31$ & $<0.34$ 90\% C.L. & --- \\
$\bdsppp$, $\dspdpi$, $\dkpipipi$ & $<1.8$ & $<2.4$ &
        $2.54\pm 0.16$ & $<0.53$ 90\% C.L. & ---\\
$\bdsppp$, $\dspdpi$, $\dkpipin$ & $<4.8$ & $<6.2$ &
        $3.18\pm 0.18$ & $<0.61$ 90\% C.L. & --- \\
$\bdsppp$, simultaneous fit & --- & --- & --- & 
        $<0.15$ 90\% C.L. & --- \\\hline\hline
  \end{tabular*}
\end{table*}
For the final result we use a simultaneous fit to the three $D^0$ decay 
channels. The $\de$ distributions for each mode were fitted together 
by a sum of a signal Gaussian and linear background function
taking into account the corresponding detection efficiencies and $D^0$ 
meson branching fractions.
The normalization of the background was allowed to float
while the signal yields were required to satisfy the constraint:
$N_i= N_{\bb}\cdot{\cal B}(\bdnpp)\cdot{\cal B}(D^0\to X_i)\cdot
\epsilon_i$,
where the branching fraction ${\cal B}(\bdnpp)$ is a fit parameter;
$N_{\bb}$ is the number of $\bb$ pairs~\cite{N_bb},
${\cal B}(D^0\to X_i)$ are the $D^0$ meson branching fractions to the 
final states $X_i$ and $\epsilon_i$ are the corresponding efficiencies.

The signals in the $\bdsnpp$ decay mode are less prominent but, when 
combined, have a $5.6\sigma$ statistical significance.
%Here, the significance is defined as 
%$\sqrt{-2ln({\cal L}_0/{\cal L}_{max})}$, where ${\cal L}_{max}$ and
%${\cal L}_0$ denote the maximum likelihood with the nominal signal
%yield and with signal yield fixed at zero, respectively.
As a cross-check, we confirm that the distribution in the 
$-0.2$~GeV$<\de<-0.15$~GeV region of the $\bdnpp$ mode
is consistent with background from $\bdsnpp$ with this measured
branching fraction.

The $\bdppp$ and $\bdsppp$ decays are doubly CKM suppressed and, thus,
are expected to have much smaller branching fractions. This is 
confirmed by the analysis of the corresponding distributions:  
we do not observe any signal for the $\bdppp$ and $\bdsppp$ decays and 
present for them upper limits at 90\% confidence level (C.L.) 
The Feldman-Cousins procedure~\cite{feldman} was used to calculate 
upper limits except for the simultaneous fit, where the maximum 
likelihood method was applied. In this case the upper limit $N$ was
calculated from the relation 
$\int^N_0 {\cal L}(n) dn=0.9\int^{\infty}_0 {\cal L}(n) dn$, 
where ${\cal L}(n)$ is the maximum likelihood with the signal 
yield at $n$. The systematic uncertainties were taken into account 
in these calculations.

\begin{figure*}
  \includegraphics[width=0.329\textwidth]{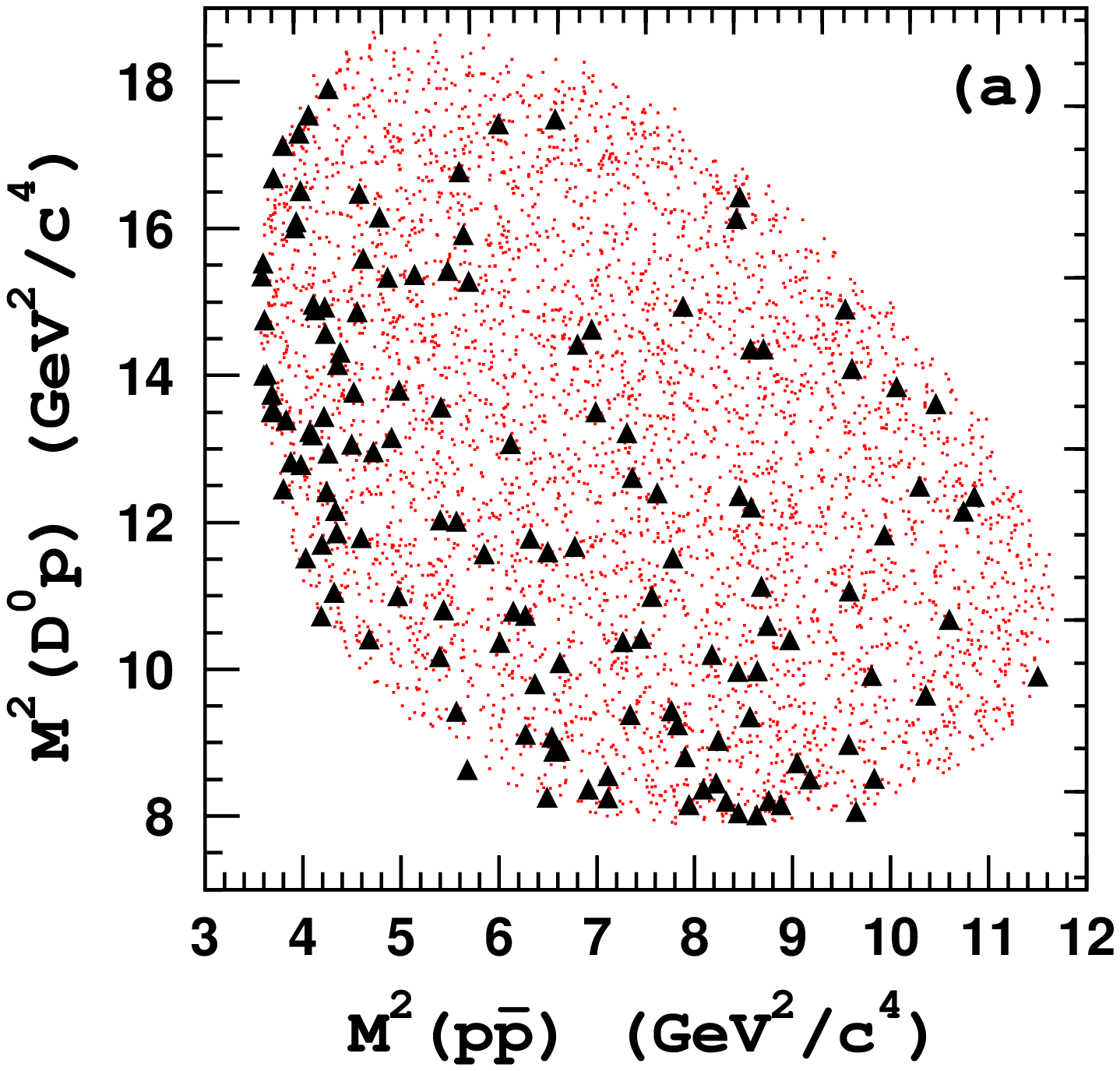}
  \includegraphics[width=0.329\textwidth]{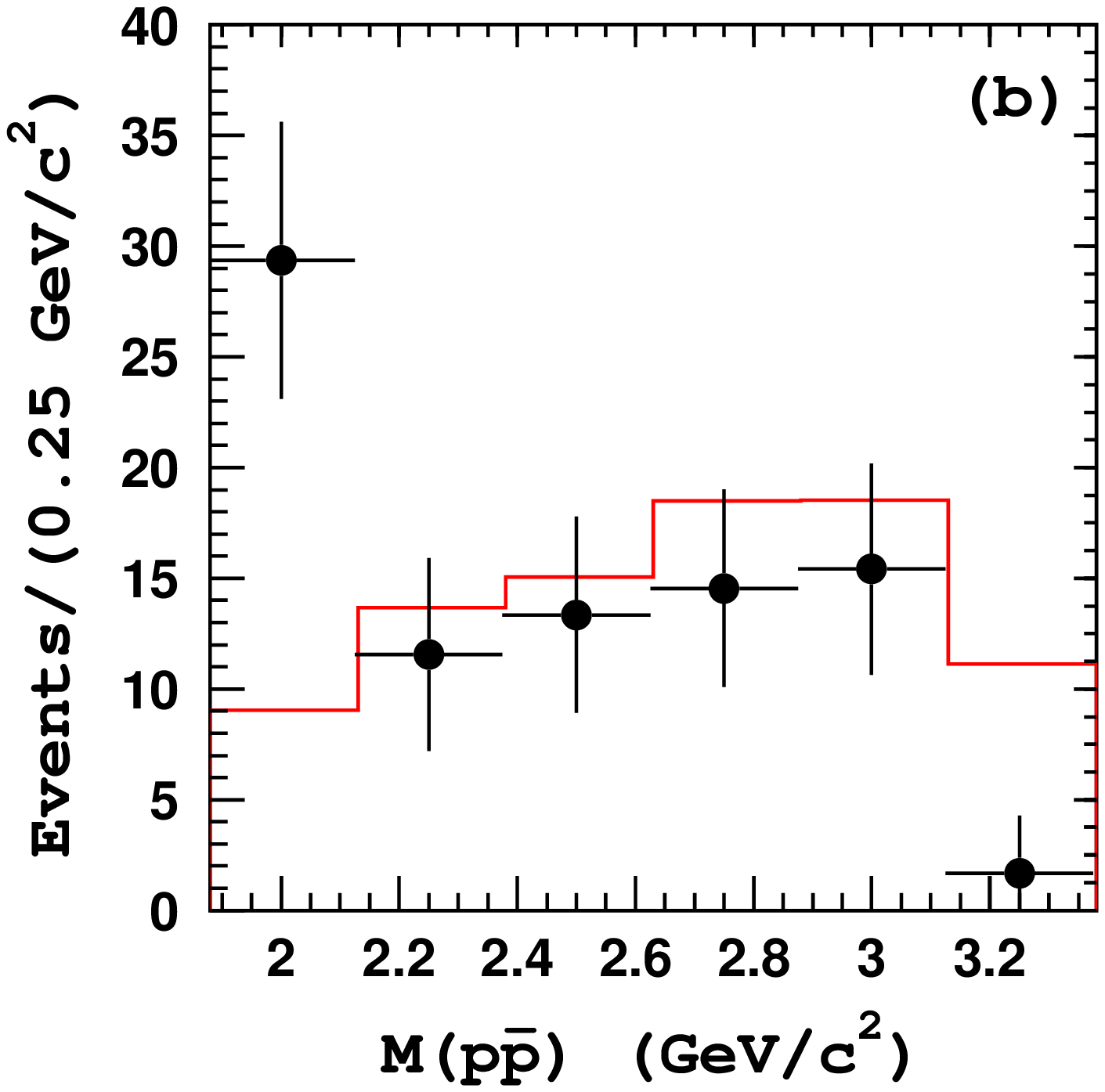}
  \includegraphics[width=0.329\textwidth]{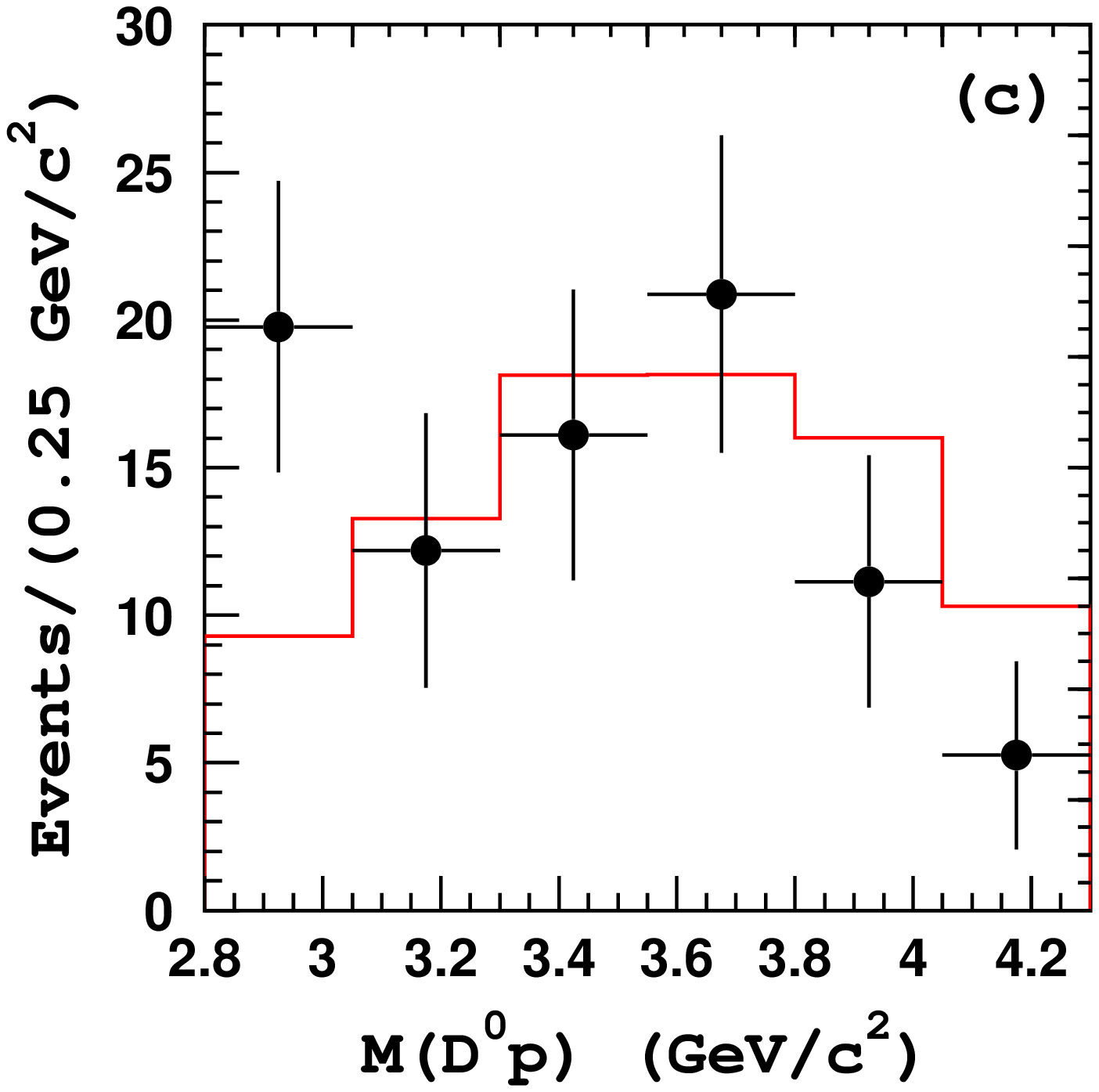}
  \caption{(a) Dalitz plot for $\bdnpp$ candidates in the $B$ 
    signal region. The triangles show events in the data and the
    small dots show the phase space simulation. The corresponding 
    invariant mass spectra 
    obtained by fitting the $\de$ distribution in each bin
    are shown in (b) for $\pp$ and (c) for 
    $D^0 p$, the data indicated by points and the phase space MC by
    histograms.}
  \label{dalitz}
\end{figure*}

%\subsection{Dalitz plot analysis}
Figure~\ref{dalitz}~(a) shows the Dalitz plot for the $\bdnpp$ candidates
from the $B$ signal region.
For comparison, also shown in Fig.~\ref{dalitz}~(a) is the same 
distribution for MC $\bdnpp$ signal events generated according to 
phase space.
It is worth noting that apart from a threshold enhancement
in the invariant mass of $\pp$ (and possibly also $D^0p$), 
the main part of the signal is distributed according to phase space.
The Dalitz plot for the $\bdsnpp$ channel (not shown) with a smaller 
statistics also reveals a similar tendency.

Since the $\pp$ invariant mass distribution of the observed signal is 
not completely described by the phase space distribution
and the detection efficiency can be non-uniform over the Dalitz plot,
some systematic uncertainty in the efficiency calculations may occur.   
To study the model dependence of the branching fractions,
we fit the $\de$ distribution for $\bdnpp$ candidates in six bins
of $\pp$ invariant mass and calculate the partial branching fraction
separately for each bin. The results are presented in
Fig.~\ref{dalitz}~(b) and in Table~\ref{mppfit}.
Summing up the partial branching fraction for each bin, we obtain the
total $\bdnpp$ branching fraction
${\cal B}(\bdnpp)=(1.12^{+0.16}_{-0.14}\pm 0.16)\times 10^{-4}$.
We apply a similar procedure to the $D^0p$ invariant mass.
The results are presented in Fig.~\ref{dalitz}~(c). 
In this case the total $\bdnpp$ branching fraction is
${\cal B}(\bdnpp)=(1.11^{+0.16}_{-0.14}\pm 0.16)\times 10^{-4}$,
consistent with the previous estimate.
The difference with the result of the simultaneous fit presented in
Table~\ref{defit} is interpreted as a model-dependent error.
The same model dependence is assumed for the $\bdsnpp$ channel.

\begin{table}
\caption{Branching fraction for the $\bdnpp$ in bins of the $\pp$ 
invariant mass.}
\smallskip
\label{mppfit}
  \begin{tabular*}{.48\textwidth}{l@{\extracolsep{\fill}}ccc} \hline\hline
$M(\pp)$, GeV & $\de$ yield & Efficiency, \%
            & ${\cal B}$, $10^{-5}$\\ \hline

$<2.13$     & $29.4^{+6.6}_{-5.9}$ & $10.01\pm 0.44$ &
        $3.65^{+0.82}_{-0.73}\pm 0.51$ \\

$2.13-2.38$ & $11.6^{+4.7}_{-4.0}$ & $9.01\pm 0.32$ & 
        $1.60^{+0.65}_{-0.55}\pm 0.22$ \\

$2.38-2.63$ & $13.4^{+4.8}_{-4.1}$ & $8.15\pm 0.28$ &
        $2.04^{+0.73}_{-0.61}\pm 0.29$ \\

$2.63-2.88$ & $14.5^{+4.8}_{-4.2}$ & $9.57\pm 0.30$ &
        $1.88^{+0.62}_{-0.54}\pm 0.26$ \\

$2.88-3.13$ & $15.4^{+5.1}_{-4.4}$ & $10.65\pm 0.34$ &
        $1.80^{+0.59}_{-0.51}\pm 0.25$ \\

$>3.13$     & $1.7^{+2.6}_{-1.7}$ & $9.18\pm 0.37$ &
        $0.22^{+0.34}_{-0.22}\pm 0.03$ \\\hline

Total  & $86.0^{+12.0}_{-10.4}$ & --- &
        $11.2^{+1.6}_{-1.4}\pm 1.6$\\\hline\hline
  \end{tabular*}
\end{table}

We examined the possibility that other $B$ meson decay modes might
produce backgrounds that peak in the signal region by means of a MC 
sample of generic $\bb$ events that corresponds to 
%%%an integrated luminosity
about 2.6 times the data statistics. No peaking backgrounds were found.

The following sources of systematic errors were found to be sizeable:
the tracking efficiency (2\% per track), proton/antiproton identification
efficiency (3\% per particle), kaon identification efficiency (2\%),
$\pi^0$ efficiency (4\%), efficiency for slow pions from $D^*\to D\pi$
decays (8\% both for $\pi^+$ and $\pi^0$), $D^{(*)}$ branching
fraction uncertainties (2\% -- 6\%),
model-dependent error (5\%) 
and MC statistics (3\% for $\bdpp$, 6\% for $\bdspp$).
The tracking efficiency error was estimated using
$\eta$ decays to $\gamma\gamma$ and $\pi^+\pi^-\pi^0$.
The proton identification uncertainty was determined from a sample of 
$\Lambda\to p\pi^-$ events; the error in kaon selection is obtained
from $D^{*+}\to D^0\pi^+$, $D^0\to \kpi$ decays.
The $\pi^0$ reconstruction uncertainty was obtained using 
$D^0$ decays to $\kpi$ and $\kpipin$.
The uncertainty in the $\de$ signal shape parameterization (3\%) was
determined by varying the mean and width of the signal Gaussian within 
their errors. 
%The model-dependent error is obtained from the $\bdnpp$
%Dalitz plot analysis, as previously mentioned.
The combined systematic error is 14\% for $\bdnpp$ ($\bdppp$) and 
17\% for $\bdsnpp$ ($\bdsppp$). 

%\section{Conclusion}
In summary, we report the first observation of the color-suppressed 
$\bdnpp$ and $\bdsnpp$ decay modes. 
The measured branching fractions are 
${\cal{B}}(\bdnpp)=(1.18\pm 0.15\pm 0.16)\times10^{-4}$ and
${\cal{B}}(\bdsnpp) =(1.20^{+0.33}_{-0.29}\pm 0.21)\times 10^{-4}$
with $12\sigma$ and $5.6\sigma$ statistical significance respectively.
No signal is observed in the $\bdppp$ and $\bdsppp$ final states. 
The corresponding upper limits at 90\% C.L. are
${\cal{B}}(\bdppp)  <0.15\times 10^{-4}$ and
${\cal{B}}(\bdsppp) <0.15\times 10^{-4}$.

We wish to thank the KEKB accelerator group for the excellent
operation of the KEKB accelerator.
We acknowledge support from the Ministry of Education,
Culture, Sports, Science, and Technology of Japan
and the Japan Society for the Promotion of Science;
the Australian Research Council
and the Australian Department of Industry, Science and Resources;
the National Science Foundation of China under contract No.~10175071;
the Department of Science and Technology of India;
the BK21 program of the Ministry of Education of Korea
and the CHEP SRC program of the Korea Science and Engineering Foundation;
the Polish State Committee for Scientific Research
under contract No.~2P03B 17017;
the Ministry of Science and Technology of the Russian Federation;
the Ministry of Education, Science and Sport of the Republic of Slovenia;
the National Science Council and the Ministry of Education of Taiwan;
and the U.S. Department of Energy.

\end{document}